\documentclass[prl,twocolumn]{revtex4}
\usepackage{graphics}
\usepackage{epsfig}

\newcommand{\twinboundaries}{TB's }

\newcommand{\muvec}{{\roarrow\mu}}

\begin{document}

\title{Multiscale nature of hysteretic phenomena: Application
to CoPt-type magnets}
\author{K. D. Belashchenko}
\author{V. P. Antropov}
\affiliation{Ames Laboratory, Ames, Iowa 50011}

\begin{abstract}
We suggest a workable approach for the description of multiscale
magnetization reversal phenomena in nanoscale magnets and apply
it to CoPt-type alloys. 
We show that their hysteretic properties are governed by two effects
originating at different length scales: a peculiar splitting
of domain walls and their strong pinning at antiphase
boundaries. We emphasize that such multiscale nature of hysteretic
phenomena is a generic feature of nanoscale magnetic materials.
\end{abstract}
\maketitle


The problem of magnetization reversal (MR) is a central and longstanding
problem in the theory of magnetism. Main complications in its treatment
stem from the fact that effects due to magnetostatic and elastic
forces manifest themselves in the whole nanoscale range while local
interactions must still be treated on the atomistic scale.
As the length scale increases, new physical effects arise, but
coarse-graining over the lower-scale degrees of freedom is not easily
accomplished. Specifically, MR is realized by motion
of domain walls (DW), the two-dimensional surfaces where 
magnetization changes its direction. DW's may easily move unless
they are pinned by defects, which implies that MR properties are
not intrinsic and depend strongly on the microstructure.
Experimentally, this results in huge variations of the coercive
force $H_{c}$
and other properties depending on material processing. It is crucial
to address the DW structure and its interaction with microstructure
on different length scales ranging from interatomic to
submicron.

In modern micromagnetic methods \cite{MM}
complex microstructures of hard magnets are defined by
(usually piecewise constant) fields of macroscopic
material properties such as the exchange constant $A({\bf r})$ and the
magnetocrystalline anisotropy (MCA) constant $K({\bf r})$.
Such calculations were mainly focussed on the role of grain
boundaries; since they are extremely hard to describe
microscopically, their properties are postulated in an
\emph{ad hoc} manner making reliable predictions problematic.
From the other hand, typical microstructures of hard magnets with
L1$_{0}$ crystal structure including CoPt, FePt and FePd
are dominated by other defects
\cite{Kandaurova-JMMM,Leroux91,Soffa2000,Khach,BPSV,Shur-CoPt},
antiphase boundaries (APB) and twin boundaries (TB), which are
crystallographically coherent and easier to
treat than grain boundaries. Although these materials were used and
studied for many decades and the formation of their microstructure
is well understood, the mechanism of MR is still a mystery.
The goal of the present letter is to lay the foundation of a
consistent multiscale theory of MR in CoPt type magnets.

Microstructural evolution during L1$_0$ ordering is strongly
affected by tetragonal lattice distortions \cite{Khach, BPSV}.
After a relatively short `tweed' stage of annealing after quench,
when the ordered domains achieve some characteristic size
$l_0\sim$~10~nm, they develop so-called polytwinning, i.e. the formation
of regular arrays (`stacks') of ordered bands (`$c$-domains') separated
by \twinboundaries~\cite{Kandaurova-JMMM,Leroux91,Soffa2000,Khach,BPSV}.
The tetragonal axes $c$ of the $c$-domains (pointing along one of
the three cubic axes of the parent fcc phase)
alternate regularly making 90$^{{\rm o}}$ angles between the adjacent
domains (below we use the obvious terms `X-domain', `XY-stack', etc.). In
addition to polytwinning, the observed and simulated microstructures always
contain a high density of APB's in the
$c$-domains~\cite{Leroux91,Soffa2000,Khach,BPSV,Shur-CoPt}.

Polytwinned magnets were extensively studied with analytical micromagnetic
methods~\cite{Kandaurova-JMMM}.
Due to high MCA the DW width
$\delta=\pi\sqrt{A/K}$ is quite small (5--10~nm), while the
anisotropy field $H_a=2K/M$ ($M$ is the saturation magnetization)
significantly exceeds the typical magnetostatic field $H_{m}=4\pi M$
(parameter $\eta=H_m/H_a$ is close to 0.1 in CoPt and FePt
and 0.38 in FePd~\cite{Kandaurova-JMMM}). Therefore, at $d\agt\delta$
($d$ is the $c$-domain thickness) each $c$-domain may be
regarded as an individual magnetic domain with intrinsic 90$^{{\rm o}}$
DW's at the \twinboundaries~\cite{Kandaurova-JMMM}. It is assumed that MR is
associated with `macrodomain walls' (MDW) crossing many
$c$-domains. Such MDW's were observed experimentally
\cite{Kandaurova-JMMM,Soffa2000}, but their internal structure
and mechanisms of coercivity are unknown~\cite{Kandaurova-JMMM}.

\paragraph{Microscopic mean-field method.}

Consider a binary alloy AB with the classical Hamiltonian 
\begin{eqnarray}
H=H_{{\rm conf}}\{n_i\}+\sum_{i<j}n_{i}n_{j}\left[ -J_{ij}{\vec{\mu}}_{i}{\vec{\mu}}%
_{j}+{\vec{\mu}}_{i}\widehat{D}_{ij}{\vec{\mu}}_{j}\right]\nonumber\\
+\sum_{i}n_{i}[\epsilon_i({\vec{\mu}}_{i})-{\bf H}_{0}{\vec{\mu}}_{i}]
\label{H-Heisenberg}
\end{eqnarray}
where $H_{{\rm conf}}$ is the configurational part of the Hamiltonian;
$i$ and $j$ run over lattice sites; $n_{i}=1$
if site $i$ is occupied by a magnetic atom A and $n_{i}=0$ otherwise
(let B be non-magnetic for simplicity); 
${\vec{\mu}}_{i}$ is the rigid classical magnetic moment of the atom at site 
$i$; $J_{ij}$, the Heisenberg exchange parameters; ${\bf H}_{0}$, the
external magnetic field; $\epsilon_i({\vec{\mu}}_{i})$, the
MCA energy equal to $-b_{i}(\muvec_{i}\mathbf{e}_i)^{2}$ for easy-axis
anisotropy; and $\widehat{D}_{ij}$, the magnetic dipole-dipole
interaction tensor.

The free energy in the mean-field approximation (MFA) may be related to the
`mean fields' ${\bf H}_{i}={\bf H}_{0}+\sum_{j}(J_{ij}-\widehat{D}_{ij})c_{j}%
{\bf m}_{j}$ where ${\bf m}_{i}=\langle {\vec{\mu}}_{i}\rangle$ and
$c_{i}=\langle n_{i}\rangle$ are local magnetizations and
concentrations, respectively:
\begin{equation}
F=-E_{J,DD}-T\sum_{i}c_{i}\ln \int d\widehat{\mu }_{i}\exp \left[%
\beta({\bf H}_{i}{\vec{\mu}}_{i}-\epsilon_i)\right].
\label{F}
\end{equation}%
Here $E_{J,DD}=\sum c_{i}c_{j}{\bf m}_{i}(-J_{ij}+\widehat{D}%
_{ij}){\bf m}_{j}$ is the total exchange and dipole-dipole energy.
Equilibrium states may be found using self-consistency relation
${\bf m}_{i}=-\partial F/\partial {\bf H}_{i}$.

If magnetization ${\bf M}({\bf r})$ slowly varies in space and is
constant in magnitude, Eq.~(\ref{F}) reduces to the
micromagnetic free energy~\cite{Aharoni}.
In this case all choices of $J_{ij}$ and $b_i$ in the defect-free
regions are equivalent if
they produce the same macroscopic properties $A$ and $K$.
However, variation of $J_{ij}$ and $b_i$ 
near defects like APB's must be studied using
first-principles techniques.
Microscopic MFA calculations with these parameters may be used
to describe DW interaction with a defect at the length scale of
$\delta$. At larger, microstructural length scales micromagnetic
methods \cite{MM} may be used
with singularities of $A$ and $K$ at the defects. However, in
hard magnets the microscopic approach also
turns out to be convenient for the studies of regions containing
up to $\sim10^6$ atoms; in such calculations some model $J_{ij}$
and $b_i$ reproducing the actual defect properties may be used.



Below, we first address the equilibrium structure of MDW's in a
defect-free system and find a peculiar MDW splitting phenomenon.
Then we explore the behavior of MDW's in external field, briefly
report our results for pinning at APB's, and finally deduce the
combined effect of MDW splitting and pinning on coercivity.

\paragraph{Equilibrium macrodomain walls.}


It is convenient to use the body-centered tetragonal (bct) representation of
the fcc lattice (with $c/a=\sqrt{2}$ and $c$ equal to the fcc lattice
parameter). Two opposite edges of a rectangular simulation box are
aligned with the (110) \twinboundaries in the XY-stack; MDW's are normal
to them. The boundary conditions are periodic, and the 
dipole-dipole fields are computed using the Fourier transforms.

\begin{figure}
\epsfig{file=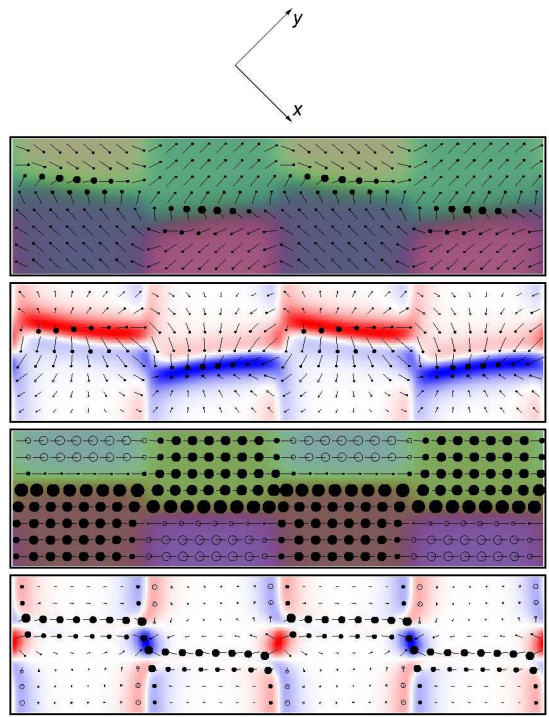,width=0.4\textwidth}
\caption{\label{DWnoAPB}Macrodomain walls in an ideal XY-stack with
$d=64a_{\rm bct}$ oriented normal
to: (a),(b) (1$\overline1$0); (c),(d) (001). Frames (a),(c):
magnetization ${\bf M}({\bf r})$ (arrows and color);
(b),(d): dipole fields (arrows) and magnetic charge density
$\rho=-{\rm div}\,{\bf M}$ (color).
The axes at top refer to (a),(b); in (c),(d) they are
rotated 90$^{\rm o}$ around the horizontal line.
Sticks with circles show vector
directions for cells $i=(8n_1+4,8n_2+4,1)$ with integer $n_1$, $n_2$.
The length of a stick (up to the center of the circle) is
proportional to the vector projection onto the graph plane;
the diameter of full (open) circles, to the
positive (negative) out-of-plane component. Small points at the end
of long sticks show their direction. Color in (a),(c)
(guide to the eye) is obtained by mixing red, green and blue
according to the values of three vector components. In (b),(d)
the intensity of red (blue) color is proportional to the
relative density of positive (negative) magnetic charges
$\rho/|\rho_{\rm max}|$. Twin boundaries
are vertical. The simulation boxes had 512x128x1 and 512x128x2
bct cells for (1$\overline1$0) and (001) MDW's, respectively.
Figures are 2 times higher than the simulation boxes; they are
trimmed at top and bottom to conserve space.}
\end{figure}

Equilibrium MDW's of two characteristic (1$\overline1$0) and (001)
orientations are shown in Fig.~\ref{DWnoAPB}
for the CoPt model. Infinite stack of $c$-domains is assumed with
ideal \twinboundaries in the (110) planes. Fully ordered
$c$-domains have the same thickness $d=64a\simeq17$~nm, which
corresponds to an early stage of
annealing shortly after the polytwinned stacks are
formed~\cite{Soffa2000,BPSV}. The anisotropy is uniaxial with 
$\mathbf{e}_i$ pointing along the local direction of the tetragonal
axis and $b_i=b$. Room temperature $T=0.4T_c$ is assumed averywhere
(for CoPt $T_c\simeq720$~K). For simplicity, only $3d$-metal atoms are
assumed to have magnetic moment $\mu$. Parameters $b$ and $\mu$
were chosen so that MFA gives experimental room-temperature values
of $K=4.9\times10^7$~erg/cm$^3$ and $\eta=0.082$ for
CoPt~\cite{Kandaurova-JMMM}. The parameters $J_{ij}$ for nearest and
next-nearest neighbors were chosen as $J_2/J_1=2/3$, $J_3/J_1=1/6$;
$A$ is isotropic with this choice.

MDW's of both orientations shown in Fig.~\ref{DWnoAPB} have a peculiar
feature which turns out to be the key to the understanding of MR in
polytwinned magnets: the DW segments (DWS) located in adjacent
$c$-domains are displaced in respect to each other. Such configuration
with regularly alternating displacements is beneficial for the exchange
energy because magnetization within each DWS is parallel to
that in the adjacent $c$-domains (note that the color of horizontal
DWS's in Fig.~\ref{DWnoAPB}a,c matches that of adjacent $c$-domains).
The magnetostatic energy for this MDW configuration is higher compared
to the rectilinear one due to the presence of short magnetically
charged (in micromagnetic terms~\cite{Aharoni}) segments
of 90$^{{\rm o}}$ DW's at the \twinboundaries (note the red and
blue `blobs' in Fig.~\ref{DWnoAPB}d). Actual MDW configuration
emerges as a result of competition between these interactions.
However, due to small $\eta$ the dipole-dipole interaction is, in
fact, unimportant at the length scale of $\delta$.

As it is clear from Fig.~\ref{DWnoAPB},
the magnetic charges both at the DWS's of a (1$\overline1$0) MDW and
at the \twinboundaries alternate in sign, so that each MDW is
magnetically uncharged as a whole, and its magnetic field quickly falls
off at $r\agt d$. DWS's of (001) MDW's do not carry magnetic
charge.

The orientation of DWS's within $c$-domains in (1$\overline1$0) MDW is
determined by the anisotropy of the exchange constant
$\alpha_A=A_\perp/A_\parallel$ ($A_\perp$ and $A_\parallel$ are values of $A$
normal and parallel to $c$), and by the parameter $\xi=\eta d/\delta$.
At $\xi\ll1$ the DW orientation is
determined solely by $\alpha_A$. At $\xi\gg1$ the DW aligns parallel to the
tetragonal axis, and $\alpha_A$ is irrelevant. The (1$\overline1$0) MDW
shown in Fig.~\ref{DWnoAPB}a,b is in the crossover region with $\xi\sim0.2$
and $\alpha_A=1$.




\paragraph{Partial macrodomain walls.}

MDW's behave remarkably in external magnetic field ${\bf H}_0$ because
DWS's are held together only by magnetostatic forces. If these
forces were absent ($\eta\to0$ limit), each DWS would be able to move freely
until it meets another DWS in an adjacent $c$-domain. Some of
this freedom remains at small $\eta$.
Consider an XY-stack with (110) \twinboundaries, as above.
Component $H_z$ does not exert any
force on DWS's, and the effect of ${\bf H}_0$ on MDW's depends only on
the orientation of its projection ${\bf H}_{\perp}$ on the $xOy$ plane.
If ${\bf H}_{\perp}$ is normal to the \twinboundaries ($H_x=H_y$), then
the forces acting on all DWS's are in the same direction, and both MDW's
shown in Fig.~\ref{DWnoAPB} move as a whole like usual DW's in a homogeneous
crystal. However, if ${\bf H}_{\perp}$ is parallel to the
\twinboundaries ($H_x=-H_y$), the forces acting on X- and Y-DWS's 
differ in sign, and the total force acting on the MDW is zero.

Let us call a set of DWS's of a MDW in all even or odd $c$-domains as
a \emph{partial macrodomain wall} (PMDW), e.g. X-PMDW.
There are two MDW realizations differing in the sign of
the relative displacement $L$ of the two PMDW's. For one of them
${\bf H}_{\perp}$ increases $|L|$ (`splitting'), and for
the other, reduces it (`swapping'). If ${\bf H}_\perp$ is
inverted, splitting turns to swapping and vice-versa.
Accordingly, two threshold fields may be introduced.
Let us fix, for instance, Y-PMDW and examine the effect
of $H_x$ on X-PMDW. 
The \emph{threshold splitting field} $H_{\text{sp}}$ is the
minimal value of $H_x$ required to move the X-PMDW
far from the Y-PMDW. The
\emph{threshold swapping field} $H_{\text{sw}}$ is 
required to `drag' the X-PMDW through Y-PMDW in the opposite
direction, overcoming the exchange barriers at the \twinboundaries.
$H_{\text{sw}}$ is of lesser importance
than $H_{\text{sp}}$ because even if
$H_{\text{sp}}<H_{\text{sw}}$, the existence of some
MDW's in the `splitting relation'
with ${\bf H}_x$ is sufficient to complete the MR. Note also that 
$H_{\text{sw}}\propto d^{-1}$ and quickly falls off with increasing $d$.
Both $H_{\text{sp}}$ and $H_{\text{sw}}$ are \emph{`intrinsic'}
properties for a given $d$, i.e.~they exist even if there are no defects
(APB's) in the $c$-domains.

\begin{figure}
\epsfig{file=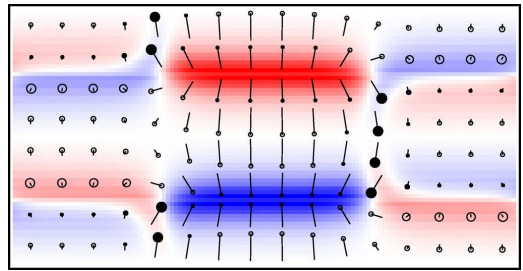,width=0.38\textwidth}
\caption{\label{splitting}Split (001) macrodomain wall in a stack with
$d=32a_{\rm bct}$.
Presentation is similar to Fig.~1d.}
\end{figure}

$H_{\text{sp}}$ may be easily estimated micromagnetically for a (001) MDW.
When its two PMDW's are displaced at distance $L$, the alternating
magnetic charges appearing at the \twinboundaries 
form a regular stack of `capacitors' (see Fig.~\ref{splitting}).
For $L\agt d$, the magnetic field is
concentrated inside these capacitors, and the total magnetostatic energy is
proportional to $L$; the coefficient in this proportionality determines
$H_{\text{sp}}$. Such calculation gives 
$H_{\text{sp}}=\pi M$ (or, alternatively, $H_{\text{sp}}=H_a\eta/4$).
The splitting threshold for (1$\overline1$0)
MDW is larger because of the magnetic charges on its segments.
In the above estimate we neglected the deviation of ${\bf M}$
from the easy axis at distances $\sim\delta$ from the \twinboundaries,
which reduces $H_{\text{sp}}$ for small $d$.

Thus, at $\eta\ll1$ we have $H_{\text{sp}}\ll H_{a}$. In particular, using the
room-temperature values of $M$~\cite{Kandaurova-JMMM},
we obtain $H_{\text{sp}}$ of 2.5~kOe
for CoPt and 3.5~kOe for FePt and FePd. These values are close to the
typical fields at which hysteresis is observed in these magnets. Quite
naturally, they are much less than the coercive fields corresponding to
uniform rotation of magnetization in the $c$-domains~\cite{Kandaurova-JMMM}.
A peculiar feature of a PMDW in the polytwinned crystal is that it separates
qualitatively different magnetic domains with different free energies, and
hence there is an intrinsic force acting on an isolated PMDW (its density per
unit cross-section of the $c$-domain is $f_{s}=2MH_{\text{sp}}$). 

\paragraph{Domain wall segments, pinning and coercivity.}

Real polytwinned alloys, as we noted above, usually contain a
high density of APB's in the $c$-domains. It is known that
DW's may be pinned by APB's~\cite{pinning}, and it was argued
that this mechanism may explain high coercivity of CoPt type
alloys~\cite{Shur-CoPt,Zhang94}. Here we show that this pinning
is indeed quite strong, but the MDW `dissociation' into segments
also has a profound effect on MR.

The APB pattern within $c$-domains determines the statistical
properties of the random value $U(x)$ where $x$ is the
coordinate of a DWS, and $U$ is its APB-induced excess
free energy. For the following basic consideration we assume
that $U(x)$ describes a distribution of similar 
pinning centers with the typical distance $l_d$ between them.
The maximal slope of $U(x)$ determines the \emph{unpinning
threshold} $H_u$, i.e. the value of the field
${\bf H}_0\parallel c$ required to unpin a single $c$-DWS.
Thus, in the absence of magnetostatic forces at $H_0<H_u$
each $c$-DWS is pinned, and at $H_0>H_u$ it can move freely
to the surface of the polytwinned stack. 
These assumptions are obvious for patterns with isolated
APB's as in Fig.~5 of Ref.~\cite{Leroux91} where
$l_d\sim5$--$10d$; however, main features of more complex
patterns~\cite{Soffa2000,Khach,BPSV,Zhang94} may also be
described by some characteristic values of $H_u$ and
$l_d\alt d$.

The maximum possible $H_{u}$ is achieved
when DWS's are parallel to isolated plain APB's.
To evaluate this maximum in the studied alloys, we explored the
modification of exchange and anisotropy at an isolated (101)-oriented APB
in CoPt, FePt and FePd using the tight-binding linear muffin-tin orbital
method. We found that MCA is strongly suppressed at the APB leading to
a DW \emph{attraction} to the APB with $H_{u}$ of about 11, 7 and 1.5~kOe
for CoPt, FePt and FePd, respectively. These values significantly
exceed the observed coercivities~\cite{Zhang94}. Since high MCA is
associated with L1$_{0}$ ordering, it is natural that local disorder
at an APB suppresses MCA.

Let us now explore the role of intrinsic magnetostatic forces in MR.
We restrict ourselves to the case of ${\bf H}_0\parallel Ox$ in
a single crystal with different types of polytwinned stacks (`single
crystal' refers to the parent fcc lattice). Such field does not affect
YZ-stacks, except for a reversible transverse magnetization. In other stacks 
${\bf H}_0$ exerts a force density $f=2MH_0$ on X-DWS's {\em only}.

From our definition of $H_{\text{sp}}$ it follows that if $c$-PMDW in X$c$-stack
($c=$Y or Z) is held
in place (e.g.,~by pinning), then the free (unpinned) X-PMDW moves to
infinity at $H_x\geq H_{\text{sp}}$ (neglecting the demagnetizing effects).
At $H_x<H_{\text{sp}}$ the X-PMDW moves to a finite distance $L(H_x)$ from
the $c$-PMDW. By definition, $L(H_{\text{sp}})=\infty$.
Except for a close vicinity of $H_{\text{sp}}$, $L\alt d$.

The path of MR differs qualitatively in two
cases: (a) $H_{\text{sp}}>H_u$, and (b) $H_{\text{sp}}<H_u$. 
In case (a) the MDW can only move as a whole (intrinsic forces $f_s$
are strong enough to unpin a PMDW), but
DWS's may to some extent `adapt' to the local pinning potential.
Let us define the characteristic displacement $l_m$
`allowed' by magnetostatic forces as $l_m=L(H_u)$. Obviously, $l_m$
increases with $H_u$, and $l_m\to\infty$ at $H_u\to H_{\text{sp}}$.
If $l_m>l_d$, the effective unpinning field of the MDW is
$H_U\sim2H_u$ (${\bf H}_0$ affects one half of the DWS's, while all
DWS's can be effectively pinned). However, as $l_m$ becomes smaller
than $l_d$, $H_U$ is quickly reduced, because only a fraction
of DWS's can be pinned simultaneously (MDW is effectively `rigid'
and can not make `kinks' comparable with $l_d$).

Now, in case (b) the MDW may split in
two PMDW's at sufficiently large ${\bf H}_0$.
The MDW is `soft', and each DWS can be effectively pinned. During the whole
cycle of remagnetization pinned Y- and Z-PMDW's do not move at all, and hence the
intrinsic (magnetostatic) force acting on a moving X-PMDW will change its
sign every time that it crosses, e.g., an Y-PMDW (in XY-stack). Thus, 
for an X-PMDW we have $H_U=H_u+H_{\text{sp}}$. Different MR properties may, in
principle, be observed if the sample was previously magnetized to saturation
by ${\bf H}_0$ along, e.g., (111). In this case there are no Y- and Z-PMDW's,
and in some stacks the switching threshold is $H_U=H_u-H_{\text{sp}}$.
The hysteresis loop in this case may
have a peculiar double-step feature with magnetization jumps at
$H_0=H_u\pm H_{\text{sp}}$. Such `hysteresis memory' effects
provide a test of the $H_{\text{sp}}<H_u$ relation and may be used
to design `programmable' magnets.

Thus, for an initially demagnetized sample with the given $H_u$ the
coercive force $H_c$ first rises with $H_{\text{sp}}$, reaches its maximum
$\sim2H_u$ at $H_{\text{sp}}\approx H_u$, stays roughly constant in the range
$\infty>l_m>l_d$, and finally falls off at $l_m<l_d$.
High coercivity of CoPt is probably due to the fact that this magnet
with suitable processing is close to the optimal $H_{\text{sp}}\sim H_u$ condition.
By contrast, in FePd $H_{\text{sp}}$ is too large ($H_{\text{sp}}/H_u\approx2.3$ even for
maximal possible $H_u$), so that $l_m\ll d$; this conclusion agrees 
with the relatively small observed $H_c/H_a$ ratio.


In our analysis we neglected the demagnetizing fields.
Close to saturation they are comparable to $H_{\text{sp}}$,
but they have little effect on $H_c$, because
they are small when the total magnetization is close to zero.


In conclusion, we developed a workable technique for the description
of multiscale MR phenomena in hard magnets and applied it to CoPt-type
alloys. We have shown that coercivity of these materials has at least
two sources originating at different length scales: strong pinning of
MDW's by APB's and their splitting at \twinboundaries. This DW
splitting seems to be a generic effect that may also affect MR in
other groups of materials, including some nanostripes~\cite{nanostripes}
and multilayers; it may also occur at grain boundaries of certain
misorientations.


\begin{acknowledgments}
The authors are grateful to V.G. Vaks for useful discussions.
This work was carried out at Ames Laboratory, which is operated for
the U.S. Department of Energy by Iowa State University under Contract
No. W-7405-82. This work was supported by the Director for Energy
Research, Office of Basic Energy Sciences of the U.S. Department of
Energy.
\end{acknowledgments}


\begin{thebibliography}{99}

\bibitem{MM} R. Fischer and H. Kronm\"uller,
Phys. Rev. B {\bf 54}, 7284 (1996); 
J. Fidler, T. Schrefl, J. Phys. D {\bf 33}, R135 (2000).

\bibitem{Kandaurova-JMMM} N.I. Vlasova, G.S. Kandaurova and N.N.
Shchegoleva, J. Magn. Magn. Mater. {\bf 222}, 138 (2000).

\bibitem{Leroux91} C. Leroux, A. Loiseau, D. Broddin and G. van
Tendeloo, Phil. Mag. B {\bf 64}, 58 (1991).

\bibitem{Soffa2000} C. Yanar, J.M.K. Wiezorek and W.A. Soffa, in
{\it Phase
Transformations and Evolution in Materials}, ed. P.~Turchi
and A.~Gonis (TMS, Warrendale, 2000), p.~39.

\bibitem{Khach} L.-Q. Chen, Y. Wang and A.G. Khachaturyan, Phil. Mag. Lett.
{\bf 65}, 15 (1992).

\bibitem{BPSV} K.D. Belashchenko, I.R. Pankratov, G.D. Samolyuk and
V.G. Vaks, J. Phys.: Condens. Matter {\bf 14}, 565 (2002).

\bibitem{Shur-CoPt} Ya.S. Shur {\it et al.}, 
Phys. Met. Metallogr. {\bf 26}, 241 (1968).

\bibitem{Zhang94} B. Zhang and W.A. Soffa, Phys. Stat. Sol. (a) {\bf 131},
707 (1992); Scr. Met. Mater. {\bf 30}, 683 (1994).

\bibitem{Aharoni} A. Aharoni, {\it Introduction to the theory of
ferromagnetism} (Clarendon, Oxford, 1996).

\bibitem{pinning} H. Kronm\"uller, J. Magn. Magn. Mater {\bf 7},
341 (1978).

\bibitem{nanostripes} M. Pratzer {\it et al.}, Phys. Rev. Lett. {\bf 87},
127201 (2001).

\end{thebibliography}
\end{document}